\documentclass[twocolumn,preprintnumbers,amsmath,amssymb]{revtex4}

\usepackage{graphicx}
\usepackage{dcolumn}
\usepackage{bm}

\newcommand{\be}{\begin{equation}}  
\newcommand{\ee}{\end{equation}}  
\newcommand{\bear}{\begin{eqnarray}}  
\newcommand{\eear}{\end{eqnarray}}  
\newcommand{\ba}{\begin{array}}  
\newcommand{\ea}{\end{array}}

\newskip\humongous \humongous=0pt plus 1000pt minus 1000pt

\newif\ifdtup

\def\oldreffmt#1{\rlap{[#1]} \hbox to 2\parindent{}}

\def\figfmt#1{\rlap{Figure {#1}} \hbox to 1in{}}  
  
\def\Tr{\mathop{\rm Tr}}

\def\slash#1{#1\!\!\!/\!\,\,}  
\def\beq{\begin{equation}}  
\def\eeq{\end{equation}}  
\def\bea{\begin{eqnarray}}  
\def\eea{\end{eqnarray}}

\def\bq{\begin{quote}}  
\def\eq{\end{quote}}

\relax  

\newdimen\tdim  
\tdim=\unitlength  
\def\bar{\overline}

\newskip\humongous \humongous=0pt plus 1000pt minus 1000pt

\newif\ifdtup

\def\oldreffmt#1{\rlap{[#1]} \hbox to 2\parindent{}}

\def\figfmt#1{\rlap{Figure {#1}} \hbox to 1in{}}  
  
\def\Tr{\mathop{\rm Tr}}

\def\slash#1{#1\!\!\!/\!\,\,}  
\def\beq{\begin{equation}}  
\def\eeq{\end{equation}}  
\def\bea{\begin{eqnarray}}  
\def\eea{\end{eqnarray}}

\def\bq{\begin{quote}}  
\def\eq{\end{quote}}

\relax  

\newdimen\tdim  
\tdim=\unitlength  
\def\bar{\overline}

\begin{document}

\preprint{FERMILAB-Pub-18-683-T}
\title{Instanton Triggered Chiral Symmetry Breaking, 
the U(1) Problem \\
and a Possible Solution to the Strong CP Problem}

\author{William A. Bardeen}

\email{bardeen@fnal.gov}

\affiliation{Fermi National Accelerator Laboratory\\
\it P.O. Box 500, Batavia, Illinois 60510, USA
}%

\date{\today}

\begin{abstract}
We argue that instantons play a crucial role in triggering the spontaneous breaking of chiral symmetry 
in the physics of the three light quarks in quantum chromodynamics. However, instantons may also 
play an essential role in our understanding of the suppression of CP violation in the strong interactions. 
We revive the possibility that the bare mass of the up quark is identically zero and the apparent small 
up quark mass is generated by the effective quark dynamics associated with instantons. In this case, 
the strong CP phase becomes unobservable and there is no strong CP violation.
\end{abstract}


\maketitle

In our present theory of the strong interactions, the six flavors of colored quarks are 
bound into color singlet bound states by the nonabelian SU(3) color gauge fields of 
quantum chromodynamics. The three light flavors of the up, down and strange quarks appear 
to generate a spectrum of eight light pseudoscalar mesons consistent with the spontaneous 
breaking of an approximate $SU(3)_L\times SU(3)_R\times U(1)$ chiral flavor symmetry. At the quark level, 
there is an additional $U_A(1)$ chiral symmetry that is seen to be badly broken by the absence of 
a ninth singlet pseudoscalar meson degenerate in mass with the light pions.

In the late 1960’s, it was discovered that some chiral symmetries are explicitly broken 
in perturbation theory \cite{One,Two,Three}. These chiral anomalies were shown to be exact 
and are not renormalized in perturbation theory \cite{Four}. In quantum chromodynamics, 
the singlet axial vector current has an anomalous divergence proportional to 
nonabelian color gauge fields. Thus, chiral anomalies have the potential of explaining 
the discrepancy observed in the spectrum of pseudoscalar mesons. However, it is precisely 
the role of instantons to provide the nonperturbative dynamics needed to realize this potential.

In this paper, we develop a new approach to how instantons trigger explicit and spontaneous breaking 
of the fermion chiral symmetries in nonabelian gauge theories.  We see that instanton dynamics 
provides a simple picture of the light meson boundstates in the Standard Model and the explicit 
breaking of the $U_A(1)$ chiral flavor symmetry.  We show how small instantons can generate 
an apparent mass for a massless up quark as a possible solution to the strong CP problem.

{\it Instantons.} The physics of instantons is intimately connected to topological structure of color gauge fields 
in the QCD vacuum \cite{Five}. Instantons are localized, nonperturbative fluctuations of the color gauge 
fields that represent quantum tunneling events \cite{Six,Seven}. In quantum field theory, a tunneling event 
can be described by a functional integral over the Euclidean gauge fields. An isolated instanton 
field configuration is described by an exact solution of the Yang-Mills field equations with one 
unit of topological charge. The weight in the functional integral is provided by the exponential 
of the Euclidean instanton action times the determinant of the fermion action.

For the instanton solution, the effective action is simply given by the inverse of the square of 
the gauge coupling constant plus a phase associated with strong CP angle, $\theta$. The phase can 
be $\pm\theta$ depending on whether the solution is an instanton solution where the gauge fields 
are self-dual or an anti-instanton solution for the anti-self-dual case:
\bea
\exp\left( -\frac{8\pi^2}{g^2}\pm i\theta \right)\det(i\slash{D}-m)
\eea
If there are no fermions, we can interpret an instanton event as a local contribution to the 
vacuum energy density of the theory in Minkowski space. The weight of the instanton contribution 
to the vacuum energy is not arbitrary but is determined by the fundamental parameters of the gauge theory. 
In a heroic calculation,  \`{}$\!$t Hooft \cite{Eight} computed the weight of the instanton contribution in an $SU(2)$ gauge theory, 
and established the precise connection between the perturbative physics of Feynman diagrams and the 
nonperturbative physics of instantons.

Instantons are highly suppressed at short distance, but their effects grow rapidly in the infrared. 
Integrating out the small instantons adds a specific correction to the effective action at larger scales. 
This correction can be represented as an integral over the parameters of the instanton including 
the size of the instanton being integrated out. For scales that are large compared to the instanton 
size, the instanton process can be viewed as a local event. All instantons with the same topological 
charge will enter with the same CP phase but will have weights that depend on the running gauge coupling 
constant as well as other factors. The result will be a local vertex whose effective coupling strength is 
determined by integrating over the instanton and anti-instanton contributions. This is usually called the 
dilute gas approximation and will have corrections if the instantons interact. For nonabelian gauge theories 
with a small number of flavors, the small instantons are highly suppressed at short distance and the dilute 
gas approximation should be very accurate. At larger scales the functional integrals may have to be done 
more precisely as the larger instantons and anti-instantons can interact and may not necessarily be viewed 
as local events.

{\it Quarks.}  When quarks are included, we must establish the impact of the instantons on the fermion determinant. 
The fermion determinant can be viewed as a product over the eigenvalues of the Dirac operator in the 
presence of the instanton field. At short distance, the modes of the free Dirac operator are bounded 
away from zero by the quantization condition on the fermion momenta. However, the instanton solution 
carries a quantized unit of topological charge. For an instanton solution, the Dirac operator of a 
massless quark has precisely one zero-mode eigenvalue for each flavor of right-handed quark and right-handed 
anti-quark. For anti-instantons, the zero-mode eigenvalues exist for the corresponding left-handed quarks 
and anti-quarks. This behavior corresponds precisely to that expected from the axial vector current anomaly 
as the divergence of the axial vector current is proportional to the topological charge density.

The presence of zero-modes for massless quarks implies that the vacuum-to-vacuum transitions will 
vanish. However, transitions where a quark and an anti-quark are created in states corresponding 
to the zero modes will not vanish. This means that the instanton transition corresponds to the 
insertion of a quark operator that causes the appropriate transitions \cite{Eight}. For a single flavor of quark, 
the required operator corresponds to quark bilinear operators, either $\{\bar{q}{}_L {q}_R\}$
or $\{\bar{q}{}_R {q}_L\}$ depending on whether 
you are dealing with an instanton or an anti-instanton. For additional massless flavors, 
effective operator will require a product of the corresponding quark fields for each flavor. 
For two flavors, this is a four-fermion operator, for three flavors this is a six-fermion operator, etc.

If we wish to determine the effects of the instanton transitions in the long distance aspects of the physics, 
we must systematically integrate out the effects of the small instantons. For massless quarks we see 
that small instantons have the effect of generating new multiquark operators. If we wish to follow a 
Wilsonian-style \cite{Nine} renormalization group procedure, we would first compute the effects of small instantons 
at a given short-distance scale which generates new effective operators or renormalizes the existing operators. 
At a larger scale, we must include the impact of the effective operators generated by the smaller instantons at 
the previous scale as well as the larger size instantons at the new scale. Hence, the effective action is 
modified by renormalizing some of the parameters of the effective action and including the new operators 
in the fermion determinant. At the larger scale, the scale of the momenta in computing the fermion determinant 
is reduced which also reduces the gap in the spectrum of the fermion modes. Integrating out the instantons at 
the new scale will require the addition of new operators corresponding to the zero-modes, or almost zero-mode 
fermions at that scale. If the new operators have the same structure as before, then we are just adding to the 
coupling strength of the previous operators. It is clear that massive fermions must also be included if their 
mass is small compared to energy scale at the distances being integrated out.

{\it No--Flavor Theory.}
In a theory without quarks, the small instantons are simply localized 
topological quantum fluctuations of the gluon vacuum which make a specific 
contribution to the vacuum energy density. The strong CP phase enters through the
relative phase of the instanton or anti-instanton contributions. At short distances, 
these fluctuations are highly suppressed by the low density of small instantons. However, 
the instanton density grows rapidly in the evolution to infrared scales and the isolated instanton 
fluctuations are replaced by more complex and possibly nonperturbative behavior of the gluon fields.

{\it One--Flavor Theory.} We now consider a theory with one quark flavor. It is clear we must make a qualitative distinction 
between massless and almost massless quarks. For massless quarks the CP phase is unobservable 
while the massive theory remains sensitive to the CP phase. For the one flavor case, the 
small instantons generate a quark bilinear operator that corresponds to an additive renormalization 
of the quark mass term. Following the renormalization group procedure, we would have to include this 
mass in the fermion determinant at the next phase of the iteration. 
The effect of this renormalization means that there are now no true zero modes. 
Instead, we have two contributions, one can be viewed as an additive shift to the 
vacuum energy, while the second is an additive shift to the effective quark mass from 
the almost zero-modes of the fermion determinant.

We now have two possibilities as we follow the renormalization group evolution. The 
first is that the perturbative gauge boson interactions do not seriously modify the 
growth of the quark mass toward the infrared. In this case, the mass grows until the gap 
in the fermion mode spectrum disappears and there are no longer any almost zero-modes 
and the quark mass is fixed. The quark is now heavy at this scale and can be integrated 
out in the further infrared evolution of the gluon dynamics as the gluons will not be 
sensitive to the now-heavy quark loops. We would also expect that when the quark mass 
is sufficiently large, the normal gluon interactions will form bound states of the 
now-heavy quarks and anti-quarks. The infrared behavior gluon fields can still affect 
the excited states. If there is no confinement, then it may be possible to see isolated 
heavy quarks. With some form of confinement, the excited states would all be bound. One could 
probe this situation by detailed studies of the heavy meson bound states.

We can illustrate this physics using a meson effective field theory. In this picture, the quark 
mass operator is replaced by a single complex meson field, $\chi$, whose phase carries the chiral 
symmetry, $\bar{q}{}_Lq_R = -Z\chi$. The usual gluon exchange dynamics is described 
by a chiral invariant effective potential for the meson fields,
\bea
V= M^2\chi^*\chi + \lambda (\chi^*\chi )^2+...
\eea
The quark mass terms are represented by tadpole operators for the bare quark 
mass and the effective mass generated by the instanton dynamics,
\bea
\Delta V= -Z(m^*\chi+\chi^*m) - G^* \chi - G \chi^*
\eea
The bare mass can have an arbitrary chiral phase while $G$ carries the
strong CP phase, $\theta$. The magnitude of $G$ varies rapidly with scale according to the
sum over the small instantons that contribute up to that scale. At short distance, $G$ is highly 
suppressed and the phase of $m$ can be rotated away. Conversely, if the 
bare quark mass vanishes the strong CP phase can also be rotated away. 
The actual phase of the effective quark mass will depend on the sum of both 
contributions at the infrared scale where the quantum fluctuations of larger instantons 
decouple from the quark physics.

An alternative picture could arise if the gluonic interactions become strong and 
nonperturbative before the quark mass stabilizes. The bound states of these lighter 
mass quarks may have a much different character than the loosely bound quark states 
of the previous picture.

{\it Two--Flavor Theory.}
We will now consider the case where there are two quarks where one quark is massless 
and the other may be massive or not. Again, as long as one of the quarks is massless, 
there is no physical consequence of the strong CP phase. At short distance, the instanton 
transitions generate a new quark operator corresponding to the zero-mode or near 
zero-mode states at the scale being considered. For two light flavors, the new quark 
operator is a four-quark interaction. It is the flavor determinant of the operators, 
$\{\bar{q}{}_{Lj}q_{Ri}\}$ for instantons  and $\{\bar{q}{}_{Rj}q_{Li}\}$ for anti-instantons,
\bea
V& = & G(\bar{q}{}_{L1}q_{R1}\bar{q}{}_{L2}q_{R2}-\bar{q}{}_{L2}q_{R1}\bar{q}{}_{L1}q_{R2})
\nonumber \\
& & +\;G^*(\bar{q}{}_{R1}q_{L1}\bar{q}{}_{R2}q_{L2}-\bar{q}{}_{R2}q_{L1}\bar{q}{}_{R1}q_{L2})
\eea
These operators must be included in the fermion determinant for the next iteration of 
the renormalization group. At short distance, the additional iterations will contribute 
as the additive renormalization of the coupling constant, $G$, with the same four quark 
operators obtained at the previous stage. The renormalization group can follow several 
paths toward the infrared physics of the two-flavor theory.

The infrared gluon dynamics can become strong and form states with light or massless quarks 
in a confinement or other phase. We can say nothing further about this possibility.

A second possibility occurs when the effective four-quark interaction becomes sufficiently 
strong to trigger the spontaneous breaking of the $SU(2)_L\times SU(2)_R \times U(1)$ chiral symmetry. 
The quarks will then develop a constituent quark mass and the now-massive quarks will 
be bound into meson states by the gluon exchange interactions. The four-quark interaction 
is attractive in the isovector channel but repulsive in the isosinglet channel for 
pseudoscalar bound states. Hence, the isovector pions will be light or massless while the isoscalar
$\eta'$ analogue will be heavy. The scalar channel
will also form bound states but the four-quark forces are reversed and the 
isovector scalars are heavy while the isoscalar scalar state is lighter 
but not expected to be massless. The nature of the highly excited states 
will depend on the infrared gluon dynamics and could be confining or not. 
The light bound states should be relatively stable against the far infrared 
behavior of the gluonic part of the theory.

A third possibility is that the heavy quark decouples and only the light, or massless, quark remains. 
In this case the four-fermion interaction induces an effective mass for the 
light quark proportional to the vacuum expectation value $\bar{q}q$ operator for the heavy quark,
\bea
m_{light}& \rightarrow & m_{light}-G\langle \bar{q}q\rangle_{heavy}  \nonumber \\
& = & m_{light}+\frac{1}{4\pi^2}GN_c\int dp^2 \frac{p^2  m_{heavy}}{p^2 + m_{heavy}^2}
\eea
This integral appears to have a quadratic divergence proportional to the bare heavy quark mass. 
However, this is fictitious as $G$ represents the infrared value of the four-fermion coupling. 
Because the instanton density decreases rapidly at short distance, the running of the effective 
coupling constant must be included and the result will be dominated by the near-infrared physics.

An intermediate case occurs when the heavy quark is still light enough for the four-quark interactions 
to trigger the spontaneous breaking of the two quark chiral symmetry in addition to explicit chiral 
symmetry breaking due to the heavy quark mass. In this case, the light quark appears to get a large 
mass from $\langle \bar{q}q \rangle_{heavy}$ if the constituent quark mass is inserted. 
However, the infrared part of this mass is already contained in physics of the spontaneous 
chiral symmetry breaking. It is only the short distance part of this expression that 
can be viewed as an effective current quark mass for the light or “massless” quark. 
Schematically,
\bea
\Delta m_{light}=\frac{1}{4\pi^2}N_c\int dp^2 G(p^2) \frac{p^2 m_{heavy,bare}}{p^2 + m_{heavy,bare}^2}
\eea
The effect may not be easily described as a simple form factor
and a more accurate treatment of the Wilsonian renormalization procedure may be required.
This mechanism shows that the instanton dynamics can generate an apparent current, 
or bare, quark mass for a quark that is massless at short distance scales. 
We will make use of this mechanism in our proposal for a solution of the strong CP problem.

{\it Three--Flavor Theory.}
As usual, nature seems to choose the most complicated case of three light flavors with a hierarchy of quark masses 
and a spontaneous breaking of the three flavor chiral symmetry, $SU(3)_L\times SU(3)_R\times U(1)$. For three flavors, the
instanton dynamics generates a six quark operator. Since the $SU(3)$ singlet pseudoscalar meson, 
the $\eta'$, is much heavier than the other pseudoscalar states, 
it must be true that instanton dynamics is essential for the observed physics of the light meson bound states.

At the quark level, the instanton dynamics generates an effective six quark operator involving 
the determinant of the quark bilinear operators, $\{\bar{q}{}_{Lj}q_{Ri}\}$ for instantons or 
$\{\bar{q}{}_{Rj}q_{Li}\}$ for anti-instantons where $\{i,j\}$ runs over the three flavors of up, down and strange quarks. 
The effective coupling constant for this operator runs rapidly with scale as it represents an integral 
over the density of small instantons that is weighted by an exponential of the inverse of the square of 
the running gauge coupling constant. We will assume that this interaction becomes strong in the infrared 
and triggers the spontaneous breaking of the three flavor chiral symmetry. In the chiral broken phase, 
the quarks develop a constituent quark mass and form bound states corresponding to the light mesons 
via the normal gluon exchange interactions. We will assume that the light meson states saturate 
correlators of the two quark operators as in chiral dynamics for the pseudoscalar mesons and vector 
meson dominance for the vector and axial vector currents. However, we now must include a nonet of 
scalar mesons and the $SU(3)$ singlet $\eta'$ meson to model the dynamics of the explicit $U_A(1)$ chiral symmetry breaking.

Using these fields we construct a simple linear sigma model to study the mechanisms of the spontaneous 
and explicit breaking of the $U(3)_L\times U(3)_R$ flavor symmetries of the naive three flavor quark theory. 
The chiral quark operators are given in terms of the meson fields by,
\bea
\{\bar{q}{}_{Lj}q_{Ri}\}=-Z\{\sigma + i \pi\}_{ij}
\eea
where $\sigma_{ij}$ and $\pi_{ij}$ represent a chiral nonet representation of meson fields. 
We use a simple effective Lagrangian 
that captures the essence of the large $N_c$ physics of the gluon exchange diagrams and the instanton physics 
incorporated in six quark, or three meson, determinant interaction.

The effective potential for the meson fields is given by,
\bea
V&=&\frac{1}{4}M^2\Tr\{\sigma^2 + \pi^2 \}
\nonumber \\
& & +\frac{1}{8}\lambda\Tr\{(\sigma-i\pi)(\sigma+i\pi)(\sigma-i\pi)(\sigma+i\pi)\}
\nonumber \\
&& -\frac{1}{2}\kappa \det(\sigma+i\pi)-\frac{1}{2}\kappa \det(\sigma-i\pi)
\eea
The determinant of the meson fields represents the instanton physics and it explicitly breaks 
the anomalous $U_A(1)$ chiral symmetry. The explicit chiral
symmetry breaking of the quark mass terms is represented by the tadpole operators,
\bea
\Delta V &=& -2m_{up}Z\{ \sigma\}_{11} -2m_{down}Z\{ \sigma\}_{22}\nonumber \\
& & -2m_{strange}Z\{ \sigma\}_{33}
\eea
In the ground state, the scalar fields will develop vacuum expectation values where $[\{ \sigma\}_{ii}
\rightarrow f_i+\{ \sigma\}_{ii}]$ as dictated by the spontaneous and explicit breaking of the chiral 
symmetries. Using this effective Lagrangian, we can fit the three neutral pseudoscalar 
states, $\pi^0,\;\eta,\; \eta'$ 
and the two decay constants, $F_K=(f+f_3)/2f$ and $F_\pi = f$.
The factor, $Z$, defines the connection between the quark operators and the 
meson fields and its value is determined using the Particle Data Group value of the strange quark mass. 
The fit determines the coupling constants for the effective field theory and the spectrum of the masses 
for all of the scalar and pseudoscalar states,
\bea
&&\!\!\!\!\!\!\!\! m_{\vec{\pi}}=135,\;\;m_\eta=548,\;\;m_{\eta'}=958,\;\;m_K=512, \nonumber \\
&&\!\!\!\!\!\!\!\! m_{\vec{\sigma}}=1110,\;\;m_{\sigma}=1360,\;\;m_{\sigma'}=740,\;\;m_\kappa=1240, \nonumber \\
&&\!\!\!\!\!\!\!\!  M=124\; \makebox{MeV}, \;\; \lambda = 35.8,\;\; \kappa=1183 \; \makebox{MeV} \nonumber \\
&&\!\!\!\!\!\!\!\!  f=\frac{1}{2}(f_1+f_2)=92\;\; \makebox{MeV}, \;\;\frac{F_K}{F_\pi}=1.19, \;\; 
\nonumber \\
&&\!\!\!\!\!\!\!\! Z= (382\; \makebox{MeV})^2, \nonumber
\eea
where the meson masses are quoted in MeV.

The fit demonstrates the nature of the $U_A(1)$ chiral symmetry breaking with the 
singlet pseudoscalar mass being enhanced and the singlet scalar mass being suppressed. 
The fit to nonet mass parameter, $M^2$, in the effective potential is small and positive providing 
phenomenological evidence that the instanton dynamics, 
not the gluon exchange interactions,  drives the spontaneous chiral symmetry breaking.

From the value of the effective coupling constant, $\kappa$,
we can read off the infrared value effective coupling for the six quark instanton operator, $G$,
\bea
V&=&-\frac{1}{2}\kappa \det(F+\sigma+i\pi)-\frac{1}{2}\kappa \det(F+\sigma-i\pi)
\nonumber \\
&=& G\det\{\bar{q}{}_{Lj}q_{Ri}\}+G\det\{\bar{q}{}_{Rj}q_{Li}\}
\eea
where $G=\kappa/(2Z^3)=(350\; \makebox{MeV})^{-5}$.

Recall that $G$ is not a conventional perturbative coupling constant 
as it is generated by integrating out the nonperturbative instanton fluctuations of
the color gauge fields. The effective value of $G$ decreases very rapidly with decreasing size, 
or higher energy scales, by its nonperturbative dependence on the exponential of the inverse 
of the gauge coupling constant,
\bea
G= \int_0^{\rho_0} d\rho \exp(-8\pi^2/g^2(1/\rho))\cdot d(\rho)
\eea
where we display the integral over the instanton size. The precise form of the the function,
$d(\rho)$ requires a detailed calculation for the specific instanton operator in three flavor QCD. 
The infrared cutoff, $\rho_0$, on the sum over instanton size would be triggered by the spontaneous 
chiral symmetry breaking and related to the scale of the constituent quark mass.

{\it The up to down quark mass ratio.}
A solution to the strong CP problem clearly resides in whether the bare up quark mass is zero. 
If we use the meson effective field theory we can estimate the value of the up 
and down quark mass splitting from the isospin splitting observed in the  $K^+- K^0$
mass difference. Correcting for the electromagnetic contributions to the charged $K$-meson mass, we use,
\bea
&& m_{K^0}= 497.6\;\makebox{MeV},\;\;m_{K^+0}= 492.4\;\makebox{MeV},\;\;
\nonumber
\eea
{where}
\bea
m^2_{K^+0}=m^2_{K^+}-m^2_{\pi^+}+m^2_{\pi^0} \nonumber
\eea
and obtain,
\bea
 m^2_{K^0}-m^2_{K^+0}=(72\; \makebox{MeV})^2
\eea
The prediction from the effective field theory is:
\bea
m^2_{K^0}-m^2_{K^+0}&=&m_\pi^2\frac{f}{f_3}\left(\frac{\lambda(2f-f_3)+2\kappa}{\lambda({f^2}/{f_3})+2\kappa}\right)
\frac{m_d-m_u}{m_d+m_u} \nonumber \\
&=& (111\;\makebox{MeV})^2\;\frac{m_d-m_u}{m_d+m_u}
\eea
From this fit, we obtain a value for the quark mass ratio, 
\bea
\frac{m_{up}}{m_{down}}= 0.42 
\eea
 which is very consistent with the current lattice value \cite{Ten}.
 
We must now determine how much of the up quark mass comes from small instanton effects 
in order to judge the evidence for a non-zero bare mass for the up quark. From the 
instanton determinant interaction, we can infer an expression for the effective up quark 
mass in terms of the vacuum expectation value of the down and strange quark densities.
\bea
m_{up } = \frac{1}{4}G\langle\bar{d}d\rangle\langle\bar{s}s\rangle
\eea
where we have assumed factorization for the multiquark operators. 
This cannot be interpreted as a bare mass term for the up quark as 
it contains the soft terms associated with the dynamical symmetry breaking. 
However, we do expect the corrections to be related to the hard down quark mass and 
instantons below the size of the meson bound states. We take the expectation value 
of the strange quark density to be its infrared value and compute the hard component 
of the down quark contribution. We have:
\bea
\Delta m_{up\;(small\;instantons) } = \frac{1}{4}G\langle\bar{s}s\rangle\langle\bar{d}d\rangle_{hard} \nonumber \\
= -\frac{1}{4}G[2Zf_3]\langle\bar{d}d\rangle_{hard}=-\frac{\langle\bar{d}d\rangle_{hard}}{(750\;\makebox{MeV})^2}
\eea
\bea
\Delta m_{down\;(small\;instantons) } = \frac{1}{4}G \langle\bar{s}s\rangle\langle\bar{u}u\rangle_{hard}
\eea
\bea
\Delta m_{strange\;(small\;instantons) } = \frac{1}{4}G \langle\bar{d}d\rangle\langle\bar{u}u\rangle_{hard}.
\eea
At a scale where the up quark is massless but the down and strange quarks
are heavy, only the up quark mass gets shifted, as $\langle\bar{u}u\rangle_{hard}$  would be zero
before the shift in the up quarks mass is included.
Using the perturbative expression for the hard down quark mass contribution, we have,
\bea
\langle\bar{d}d\rangle_{hard}=-\frac{4N_c}{16\pi^2}\int dp^2\; m_{d,\; hard}
\eea
The integral is apparently quadratically divergent at high momenta. 
However, at high momenta the instanton effective coupling, $G$, decreases very rapidly so 
the integral should only have contributions from above the soft symmetry breaking 
scale to a scale associated with a short distance cutoff reflecting the absence of small instantons. 
Using our values for the effective field theory parameters, we obtain the estimate,
\bea
\Delta m_{up } = \frac{(\Lambda^2-\mu^2)}{(2.72\;\makebox{GeV})^2} m_{d,hard}
\eea
where $\Lambda$ is an estimate of scale for the ultraviolet suppression of small 
instantons and $\mu$ is the cutoff scale associated with the soft physics.
Using $\Lambda=2$ GeV and $\mu = 1$ GeV we obtain an estimate of the instanton generated 
current quark mass for the up quark,
\bea
\frac{m_{up}}{m_{down}} = 0.40
\eea
which is very consistent with the value required to fit the data within the framework of the soft 
dynamics alone. We conclude that small instanton effects can plausibly generate the observed 
up quark mass without the need for a bare mass for the up quark.

The massless up quark solution to the strong CP problem is again viable. 
While we have a qualitative picture of how instanton induced operators are generated 
in a nonperturbative treatment of quantum chromodynamics, it is difficult to see how 
to make this very quantitative. The solution to the $U_A(1)$ problem requires a large effective 
coupling for the instanton induced determinant operator in the far infrared which should imply 
that it will have
important effects at nearby scales as needed to generate the effective up quark mass 
in the low energy fits to the data.

{\it Conclusions.}
We have argued that the nonperturbative effects of instantons in quantum chromodynamics can be 
systematically included using a Wilsonian-like renormalization group picture where the small instantons 
are integrated out and replaced by effective quark operators. These effective quark operators 
must then be included in the fermion determinant when considering larger scales. The contributions 
of the larger instantons generate the same effective quark operators as the smaller instantons 
but with prefactors that rapidly increase as we approach the infrared scales. At a given scale, 
the effective coupling constant for the instanton induced operator is given by the sum over the 
contributions from instantons at all scales below the scale being considered. Because small instantons 
are highly suppressed, the effective coupling constants for the instanton induced operators will also 
be highly suppressed at short distance scales but will increase rapidly until they generate a nonperturbative 
infrared scale. This can happen by explicitly generating an effective mass for the quarks, as in 
the one flavor theory or by triggering the dynamical breaking of chiral symmetries of light quarks. 
The quarks will then decouple from the soft physics of the gluon fields. 

One may worry that the 
effective operators induced in multi-flavor theories are nonrenormalizable in the conventional sense. 
However, they remain multiplicatively renormalizable in massless perturbation theory as they
are the unique operators of maximal chirality.
The strong suppression of instantons at short distance means that the main renormalization 
effects are associated with the running of the QCD coupling constant in the exponent of the prefactor 
associated with the instanton contribution to the partition function. This picture could change 
if the theory has a large number of flavors as the running of the effective coupling constants 
for instanton induced operators are much slower and renormalization effects or the interaction 
between instantons at different scales may become important.
Of course, soft gluon physics can reappear when studying the excited states of the now heavy quarks. 
Issues like the nature of confinement in the far infrared and the creation of QCD strings 
are not addressed in this study.

Finally, we comment on the nature of scale invariance in the presence of instantons. 
The perturbative physics of QCD is governed by the logarithmic running of the QCD coupling 
constant and above the scales of one or two GeV the breaking of scale symmetry appears to be 
suppressed and is described in terms of the QCD beta-function. However, in the instanton sector, 
scale symmetry is badly violated by the running of effective coupling constant associated with the 
higher dimensional multiquark operators generated by quark zero-modes of quantum tunneling 
transitions induced by instanton fluctuations of the gluon fields. In our applications to quantum 
chromodynamics, this dramatic violation of scale symmetry is not visible at
short distance as the small instantons are highly suppressed. However, the same strong running 
insures that the instanton induced operators will play an essential role in determining the 
infrared structure of the QCD. Indeed, we see in our simple effective field theory describing 
the physics of three flavor QCD that instanton effects appear to drive the spontaneous breaking 
of the light quark chiral symmetries that establishes the mass scale for the spectrum of meson and 
baryon states that we see in nature.

We have seen that the physical scale of the infrared physics can be determined by the interplay 
between the positive “perturbative” dynamics scaling like powers of the gauge coupling constant 
and the negative, exponentially suppressed instanton terms. In our examples of the one quark 
flavor case and the three quark flavor case with massless bare quarks, an intermediate scale 
emerges in a massless theory by a form of dimensional transmutation. The solution does not require 
the quarks and gluons to appear to have nonperturbative interactions in the usual sense. Perhaps all 
scales are determined by the interplay between the “perturbative” gauge boson dynamics and a variety of 
instantons in various sectors of a complete theory.

\vskip .1in
\noindent
{\it Acknowledgements}.
I would like to thank my colleagues, Estia Eichten, Chris Hill, Ciaran Hughes, William Jay, 
and Chris Quigg for useful comments and discussions throughout the course of
this study. This work was supported by the Fermi Research Alliance, LLC, under Contract 
No. DE-AC02-07CH11359 with the Department of Energy, Office of Science, Office of High Energy Physics.

\end{document}